\documentclass[fp,twocolumn]{jpsj3}
\usepackage{txfonts}

\title{Microscopic Observation of Heavy Quasiparticle Formation in the Intermediate Valence Compound EuNi$_2$P$_2$: $^{31}$P NMR Study}

\author{
\name{Nonoka Higa}$^1$,
\name{Mamoru Yogi}$^2$\thanks{myogi@sci.u-ryukyu.ac.jp},
\name{Hiroko Kuroshima}$^1$,
\name{Tatsuki Toji}$^1$,
\name{Haruo Niki}$^2$,
\name{Yuichi Hiranaka}$^1$,
\name{Ai Nakamura}$^{1}$,
\name{Takao Nakama}$^2$,
\name{Masato Hedo}$^2$, and
\name{Yoshichika \={O}nuki}$^2$
}

\inst{
$^1$Graduate School of Engineering and Science, University of the Ryukyus, Nishihara, Okinawa 903-0213, Japan\\
$^2$Faculty of Science, University of the Ryukyus, Nishihara, Okinawa 903-0213, Japan
} %\\

\abst{
We report $^{31}$P NMR measurements under various magnetic fields up to 7 T for the intermediate valence compound EuNi$_2$P$_2$, which shows heavy electronic states at low temperatures.
In the high-temperature region above 40 K, the Knight shift followed the Curie--Weiss law reflecting localized $4f$ states.
In addition, the behavior corresponding to the temperature variation of the average valence of Eu was observed in the nuclear spin-lattice relaxation rate $1/T_1$.
With the occurrence of the Kondo effect, $1/T_1$ was clearly reduced below 40 K, and the Knight shift becomes almost constant at low temperatures.
From these results, the formation of heavy quasiparticles by the hybridization of Eu $4f$ electrons and conduction electrons was clarified from microscopic viewpoints.
Furthermore, a characteristic spin fluctuation was observed at low temperatures, which would be associated with valence fluctuations caused by the intermediate valence state of EuNi$_2$P$_2$.

}

\begin{document}
\maketitle

\section{Introduction}

The localization and itinerancy of the $f$ electron in rare-earth compounds are closely related to their ground states.
The Kondo effect and the Ruderman--Kittel--Kasuya--Yosida (RKKY) interaction are important interactions between $f$ electrons and conduction electrons, and the elucidation of specific physical properties in the vicinity of the quantum critical point (QCP), where these interactions are competing, is one of the essential themes in the strongly correlated electron system.\cite{Lohneysen2007,Gegenwart2008}
After the discovery of the well-known first heavy electron superconductor CeCu$_2$Si$_2$,\cite{Steglich1979} many studies have been conducted on peculiar physical properties near the antiferromagnetic QCP.\cite{Lohneysen2007,Gegenwart2008}
Superconductivity was also found in the ferromagnetic materials UGe$_2$, URhGe, and UCoGe.\cite{Saxena2000,Aoki2001,Levy2005,Huy2007}
The electronic states near the ferromagnetic QCP with pressure application and the magnetic field have also attracted considerable attention, and many research studies have been carried out.

%希土類化合物におけるf電子の局在性と遍歴性はそれらの基底状態と密接に関連している。
%重要な相互作用として近藤効果やRKKY相互作用があり、そしてそれらの拮抗した量子臨界点近傍における異常物性の解明は強相関電子系における重要なテーマの一つである
%よく知られた初の重い電子超伝導体CeCu2Si2の発見後、反強磁性量子臨界点近傍における異常物性について多くの研究が行われてきた。
%また、強磁性体UGe2やURhGe、UCoGeなどにおいても超伝導が見いだされた。
%圧力印可や磁場印可による強磁性臨界点近傍の電子状態についても大きな注目を集め、多くの研究が進められてきた。

The relationship between the valence states of $f$ electrons and novel electronic states is also being studied.
For example, it has been reported that the superconducting transition temperature $T_c$ of CeCu$_2$Si$_2$ is greatly enhanced by applying pressure.\cite{Yuan2003,Holmes2004}
The crossover of the valence of Ce is experimentally observed at around 4 GPa where $T_c$ is maximum,\cite{Kobayashi2013} and the occurrence of superconductivity due to valence fluctuations is theoretically discussed.\cite{Miyake2014}
In addition, a heavy electronic state in Yb compounds has been found in YbCo$_2$Zn$_{20}$ and $\beta$-YbAlB$_4$.\cite{Torikachvili2007,Nakatsuji2008}
It has been reported that $\beta$-YbAlB$_4$ shows not only a heavy electronic state but also quantum criticality different from that of the Doniach diagram, which is often observed in Ce compounds.\cite{Tomita2015}
The valence of Yb in $\beta$-YbAlB$_4$ is estimated to be +2.75, and the intermediate valence (mixed-valence) state is presumably related to the novel ground state of $\beta$-YbAlB$_4$.\cite{Okawa2010}
Thus, the valence instability of rare-earth ions plays an important role in the emergence of novel physical properties.
%f電子の価数状態と新奇な電子状態の関連についても研究が進められている。
%例えば、CeCu2Si2の超伝導転移温度Tcは、圧力印加により大幅にエンハンスする事が報告されている。
%Tcが最大となる4GPa近傍において、Ceの価数のクロスオーバーが実験的に示唆されており、理論的にも価数揺らぎによる超伝導の実現について議論されている。
%また、Yb化合物においても重い電子状態が実現することがYbCo2Zn20やβ-YbAlB4などにおいて報告されている。
%β-YbAlB4は重い電子状態のみならず、超伝導やドニアックピクチャーとは異なる量子臨界性を示すことが報告されている。
%また、β-YbAlB4においてYb価数は+2.75と見積もられており、中間価数状態にある事がこれらの新奇物性と関連していると推測される。
%よって、希土類元素化合物における価数のinstabilityは新奇物性発現に重要な側面を持っていると考えられる。

In this study, we focus on the rare-earth element Eu, which has two kinds of valence states Eu$^{2+}$ ($4f^7$) and Eu$^{3+}$ ($4f^6$).
The divalent Eu state is magnetic ($J=S=7/2$, $L=0$), where $J$ is the total angular momentum, $S$ is the spin angular momentum, and $L$ is the orbital angular momentum.
Therefore, intermetallic compounds with divalent Eu ions tend to order magnetically at low temperatures, following the RKKY interaction.\cite{Felner1975,Felner1978}
The pressure effect on divalent Eu compounds is often different from that of Ce compounds.
For example, in EuNi$_2$Ge$_2$ and EuRh$_2$Si$_2$, a valence transition from a divalent state to an almost trivalent state occurs above the critical pressure $P_V$, and a nonmagnetic ground state is realized.\cite{Wada1999,Nakamura2012,Mitsuda2012,Honda2016}
On the other hand, some compounds, such as EuPt$_{2}$Si$_{2}$, behave following the Doniach diagram.
The antiferromagnetic transition temperature $T_{\rm N}$ of EuPt$_2$Si$_2$ decreases continuously with increasing pressure without showing the valence transition, and $T_{\rm N}=0$ at $P_V \simeq  4$ GPa.\cite{Mitsuda2003,Takeuchi2018}
%本研究では希土類元素Euに着目する。
%一般にEuは磁性を持つ2価と非磁性の3価の状態を取る。
%化合物中において、Euの価数は2価になる場合が多く、低温で磁気秩序状態が実現する。
%圧力印可による電子状態の変化は、Ce系化合物とは異なる場合が多く、臨界圧力PV以上で価数転移を示すことがEuNi2Ge2やEuRh2Si2等において報告されている。
%一方で、ドニアック描像に従うような物質も存在し、EuPt2Si2では圧力増加に伴い反強磁性転移温度$T_{\rm N}$が減少することが報告されている。
Unlike the divalent Eu state, the trivalent Eu state is nonmagnetic ($J=0$, $S=L=3$).
EuPd$_3$ is one of the few Eu compounds in which the Eu ion becomes trivalent.\cite{Harris1965,Harris1971,Cho1996}
The specific heat and magnetic susceptibility of EuPd$_3$ are analyzed using the $J$-multiplet levels $^{7}F_{J}$ with $J=0-6$.\cite{Gardner1972,Takeuchi2014}
%二価のEu状態とは違い、三価のEu状態は非磁性となる。
%非磁性基底状態をとる化合物としてEuPd3があり、それらの価数は3価である事が実験的に示されている。

Among Eu compounds, EuNi$_2$P$_2$ takes an intermediate valence state and shows a significant temperature variation in the average valence.
It has been reported from M\"{o}ssbauer measurements that the average valence of Eu is +2.25 at 300 K, and it increases with decreasing temperature and becomes +2.50 at 1.4 K.\cite{Nagarajan1985}
Specific heat measurements down to 80 mK have revealed that EuNi$_2$P$_2$ shows no magnetic order and forms a heavy electronic state with a large electronic specific heat coefficient $\gamma = 93$ mJ/(K$^2$$\cdot$mol) at low temperatures.\cite{Fisher1995,Hiranaka2013}
For the trivalent compound EuPd$_3$, the electronic specific heat coefficient was obtained as $\gamma = 3.6$ mJ/(K$^2$$\cdot$mol).\cite{Takeuchi2014}
Therefore, the intermediate valence state of Eu plays an important role in the formation of the heavy electronic state in EuNi$_2$P$_2$.
The hybridization between the Eu $4f$ electrons and conduction electrons was reported from studies of photoemission spectroscopy, X-ray magnetic circular dichroism, and optical conductivity.\cite{Danzenbacher2009,Matsuda2009,Guritanu2012,Anzai2017}
In these circumstances, we carried out $^{31}$P NMR measurements on EuNi$_2$P$_2$ to clarify its electronic state from a microscopic viewpoint.
We report on the static magnetic properties and low-energy fluctuations of EuNi$_2$P$_2$ from the measurements of resonance spectra and relaxation rates under various magnetic fields up to 7 T.
%その中で、EuNi2P2は中間価数状態を取る化合物である。
%室温で○価であり、低温では○価となることがメスバウアーの測定から報告されている。
%EuNi2P2の特徴は磁気秩序を示さず、大きな電子比熱係数γを有する重い電子状態を低温で形成することである。
%Euが3価であるEuPd3の電子比熱係数はBBBとなり大きくない。
%よって、EuNi2P2の重い電子状態形成にはEuが中間価数状態である事が重要であると考えられる。
%以上の背景の下、我々は、EuNi2P2の価数の変化や重い電子状態の形成について微視的な視点から明らかにするために、P-NMRによる研究を行った。
%本論文では、様々な磁場下におけるナイトシフトと緩和時間に関する詳細な測定結果について報告を行う。

\section{Experimental Procedure}
Single crystals of EuNi$_{2}$P$_{2}$ were grown by the Sn-flux method.
Details of the sample preparation are described elsewhere.\cite{Marchand1978,Hiranaka2013}
The crystals were powdered to facilitate applied rf-field penetration.
The NMR measurement was performed on $^{31}$P nucleus (nuclear spin $I=1/2$) by a conventional spin-echo method using a conventional phase-coherent pulsed spectrometer in the temperature range of $T = 1.6-300$ K.
The NMR spectra were obtained by sweeping the frequency and integrating the spin-echo signal intensity step by step.
The Knight shift was referred to the $^{31}$P resonance frequency of phosphoric acid solution, $\nu_0 \equiv  (^{31}\gamma_n/2\pi) \mu_0H$.
Here, $^{31}\gamma_n$ is the nuclear gyromagnetic ratio of $^{31}$P and $\mu_0H$ is an external magnetic field.
The nuclear spin-lattice relaxation time $T_1$ was measured by a saturation-recovery method.

\section{Results and Analysis}
\subsection{NMR spectrum and Knight shift}

\begin{figure}[tb]
  \begin{center}
    \includegraphics[keepaspectratio=true,width=80mm]{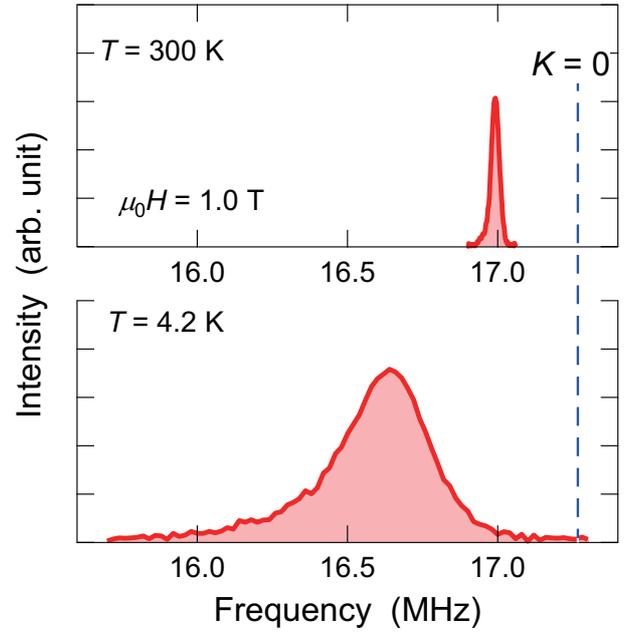}
  \end{center}
  \caption{(Color online) $^{31}$P NMR spectra of EuNi$_{2}$P$_{2}$ in $\mu_0H=1.0$ T at $T =$ 300 and 4.2 K.}
  \label{fig1}
\end{figure}
Figure 1 shows the $^{31}$P NMR spectra of EuNi$_2$P$_2$ measured in an external magnetic field $\mu_0H = 1.0$ T at temperatures $T = 300$ and 4.2 K. 
At 300 K, a sharp spectrum with a line width of 190 kHz was observed.
With decreasing temperature, the line width increases, and the peak position shifts to the low-frequency side.
Since $^{31}$P has no nuclear quadrupole moment, the resonance frequency $\nu_{m}$ can be described as
%図1にEuNi2P2のP-NMRスペクトルを示す。
%300Kでは半値幅000kHzのシャープな1本のスペクトルが観測された。
%温度降下に従い、線幅は磁化率と同様に増大し、ピーク位置は低周波にシフトしていった。
%31Pは核スピン$I=1/2$なので、核四重極モーメントを有しておらず、磁気的な相互作用のみが存在する。
%よって、共鳴周波数$\nu_{m}$は以下のようにかける。
\begin{equation}
\nu_{m} = \nu_0 \left[1+K(\theta)\right].
\end{equation}
Here, $K(\theta)$ is the Knight shift, and $\theta$ is the angle between the external magnetic field and the $c$-axis of the crystal.
From the local symmetry of the P site in EuNi$_2$P$_2$ ($4mm$ in Hermann-Mauguin notation or 
international notation), the Knight shift has axisymmetric anisotropy.
In this case, the Knight shift can be written as $K(\theta) = K_{\rm iso} + K_{\rm an}(3\cos^2\theta - 1)$ using the isotropic shift $K_{\rm iso}$ and the anisotropic shift $K_{\rm an}$.
The characteristic powder patterns should be observed when we use randomly oriented powder samples.
The shape of the obtained spectrum indicates no remarkable anisotropy in the Knight shift.
The spectrum at 4.2 K shows a tail to the low-frequency side, which possibly corresponds to anisotropy appearing in magnetic susceptibility below about 50 K.\cite{Hiranaka2013}
%ここで、$K(\theta)$はナイトシフトで、$\theta$は外部磁場と$c$軸のなす角である。
%EuNi2P2は正方晶構造をとるため、ナイトシフトは等方的なシフト$K_{iso}$、異方的なシフト$K_{an}$を用いて$K(\theta) = K_{iso} + K_{an}(3\cos^2\theta - 1)$とかける。
%よって、ランダムに配向した試料を用いた場合、特徴的な粉末パターンが観測される。
%観測された共鳴スペクトルの形状から、ナイトシフトには顕著な異方性が無いことがわかる。
%4.2Kのスペクトルはわずかに低周波側に裾を引いているように見える。
%これは約50Kでは磁化率にも異方性が現れることに対応していると推測される。

\begin{figure}[tb]
  \begin{center}
    \includegraphics[keepaspectratio=true,width=80mm]{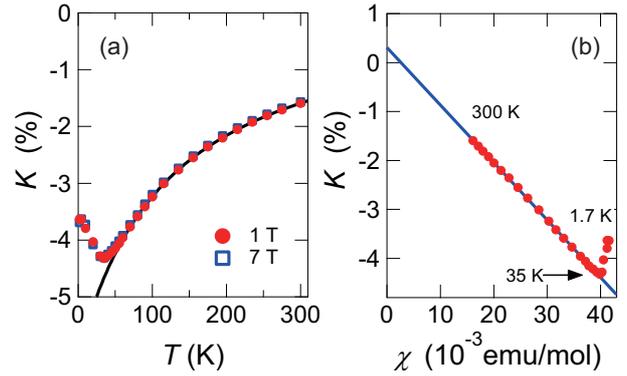}
  \end{center}
  \caption{(Color online) (a) Temperature dependences of the Knight shift at $\mu_0H=1.0$ and 7.0 T. The solid curve indicates a Curie--Weiss fit. (b) $K$ vs $\chi$ plot with $T$ as implicit parameter. The solid line indicates $K=[A_{\rm hf}/(N_{\rm A}\mu_{\rm B})]\chi$ with hyperfine coupling constant $A_{\rm hf}=-6.53$ kOe/$\mu_{\rm B}$.}
  \label{fig2}
\end{figure}

Since the anisotropy of the spectrum was small as described above, the Knight shift was determined by the peak position of the spectrum.
The temperature dependences of the Knight shift measured under the magnetic fields $\mu_0H = 1.0$ and 7.0 T are shown in Fig. 2(a).
The obtained Knight shift at $\mu_0H =1.0$ T is in agreement with the previously reported results \cite{Sampathkumaran1985,Magishi2015} and shows no significant magnetic field dependence up to $\mu_0H = 7.0$ T.
This indicates that the static magnetic and valence states of Eu are not affected by the magnetic field.
In addition, the Knight shift has a negative value in the whole measurement temperature range, indicating that the core polarization due to $2p$ electrons is a major contribution to the Knight shift.
%上記のように、スペクトルの異方性は小さいと考えられるため、スペクトルのピーク位置からナイトシフトを決めた。
%外部磁場1Tおよび7Tの下で測定したナイトシフトの温度依存性を図2(a)に示す。
%低磁場におけるナイトシフトはこれまでに測定されており、今回の測定結果は同様であった。
%また、最大7Tの外部磁場でナイトシフトを求めたが、磁場依存性は見られなかった。
%これは、Euの静的な磁気状態および価数状態について、磁場による影響が無い事を示している。
%また、すべての温度領域でナイトシフトは負の値を取った。
%これはPの2p電子によるコア偏極による寄与が主要である事を示している（Euの4f電子が2pにトランスファーでとか書く？）。
In general, the Knight shift consists of the sum of the spin part $K_{\rm s}$ and the orbital part $K_{\rm orb}$.
Here, assuming $K_{\rm s}$ obeys the Curie--Weiss law and $K_{\rm orb}$ is independent of temperature, the Knight shift can be written as
\begin{equation}
K(T) = \frac{C}{T - \theta} + K_{\rm orb}.
\end{equation}
The solid line in Fig. 2(a) is the result of the fitting for the experimental data above 100 K with $C$, $\theta$ and $K_{\rm orb}$ being the free parameters.
We obtained $C=-768$ \%K, $\theta = -121.8$ K, and $K_{\rm orb} = 0.23$\%.
The fitted line is in good agreement with the experimental data, which means that the static magnetic properties are dominated by the localized moments of the $4f$ electrons in Eu at the high-temperature region.
%Since a divalent state of Eu has no total orbital angular momentum ($L = 0$), observation of finite $K_{\rm orb}$ is presumed to be because EuNi$_2$P$_2$ is in the intermediate valence state.
A divalent state of Eu has no total orbital angular momentum ($L = 0$).
Therefore, the finite $K_{\rm orb}$ is presumed to be due to the intermediate valence state of Eu in EuNi$_2$P$_2$.
%This result reveals that the static magnetic properties are dominated by the localized moments of the $4f$ electrons in Eu at high-temperature region.
%Eu$^{2+}$ state has $L=0$; therefore, the observation of finite $K_{\rm orb}$ is presumably due to an intermediate valence state of EuNi$_2$P$_2$.
%The fitted line is in good agreement with the experimental data, which means that the static magnetic properties are dominated by the localized moments of the $4f$ electrons in Eu at high-temperature region and the valence state of Eu in EuNi2P2 is not divalent but the intermediate state because of $Korb \neq 0$.
%On the other hand, even though Eu$^{2+}$ has $L = 0$, finite $K_{\rm orb}$ was also observed.
%This is considered to be due to the intermediate valence state of Eu in EuNi$_2$P$_2$.
%通常、ナイトシフトはスピンパート$K_{\rm s}(T)$と軌道パート$K_{\rm orb}$の和からなる。
%ここで、Ksがキュリーワイス則に従い、Korbは温度に依存しないとすると、以下のように書くことができる。
%図2(a)の実線は100K以上の実験結果について式でフィットした結果であり、C=-768%、θ=-121.8、Korb=0.23%が得られた。
%この結果から、ナイトシフトはEuの4f電子の局在モーメントにより決まっていることが示される。
%一方でEu$^{2+}$では$L=0$である事に反して、有限の$K_{\rm orb}$が観測された。
%これはEuNi2P2ではEuが中間価数状態を取ることに起因していると推測される。
%100K以下では、ナイトシフトはCurie-Weiss則からずれ始め、40Kでその絶対値は最大となり、その後、減少に転じる振る舞いが観測された。
%これは、近藤効果により局在モーメントが伝導電子と混成することにより、重い準粒子が形成され、パウリ常磁性的な振る舞いへと変化した物だと考えられる。

At low temperatures below about 70 K, the Knight shift deviates from the Curie--Weiss behavior and shows a minimum at 40 K.
At the lowest temperature, the Knight shift becomes almost constant.
%At low temperatures, Pauli paramagnetic behavior was observed.
These characteristic behaviors indicate the formation of a heavy electron state due to the Kondo effect.
Figure 2(b) shows the $K$ vs $\chi$ plot.
There is a good linearity above 35 K; however, a change in the slope is observed at lower temperatures.
Such behavior was observed in many heavy-fermion compounds,\cite{Curro2004} the origin of which is due to the change in the hyperfine coupling constant associated with heavy quasiparticle formation.
Thus, the $K-\chi$ plot suggests the formation of a heavy electron state at low temperatures.
%KvschiプロットをFig.2(b)に示す。
%35K以上の温度領域では、よい直線性を示している。
%ナイトシフトの絶対値が最大となる約40K以下で、K-chiプロットは折れ曲がっている。
%この様な振る舞いは多くの重い電子系化合物で観測されており、その起源は重い準粒子形成に伴う超微細結合定数の変化による。
%よって、K-chiプロットの様子からも、低温で重い電子状態の形成が示唆される。
The relationship between the hyperfine coupling constant of $A_{\rm hf}$ of $^{31}$P and the magnetic susceptibility $\chi$ can be expressed as 
\begin{equation}
K = \frac{A_{\rm hf}}{N_{\rm A}\mu_{\rm B}}\chi,
\end{equation}
where $N_{\rm A}$ is Avogadro's number and $\mu_{\rm B}$ is the Bohr magneton.
The hyperfine coupling constant of $A_{\rm hf}=-6.53$ kOe/$\mu_{\rm B}$ is estimated from the slope of the solid line, which is fitted to the experimental data above 30 K.
%The hyperfine coupling constant $A_{\rm hf}$ of $^{31}$P can be estimated from the slope of $K-\chi$ plot with the relation
%\begin{equation}
%K = \frac{A_{\rm hf}}{N_{\rm A}\mu_{\rm B}}\chi,
%\end{equation}
%where $N_{\rm A}$ is Avogadro's number, and $\mu_{\rm B}$ is the Bohr magneton.
%The solid line in Fig. 2 (b) is a fit of data above 35 K, and the hyperfine coupling constant is estimated to be $A_{\rm hf}=-6.53$ kOe/$\mu_{\rm B}$.

\subsection{Analysis of $T_1$}

\begin{fullfigure}[t]
  \begin{center}
    \includegraphics[keepaspectratio=true,width=160mm]{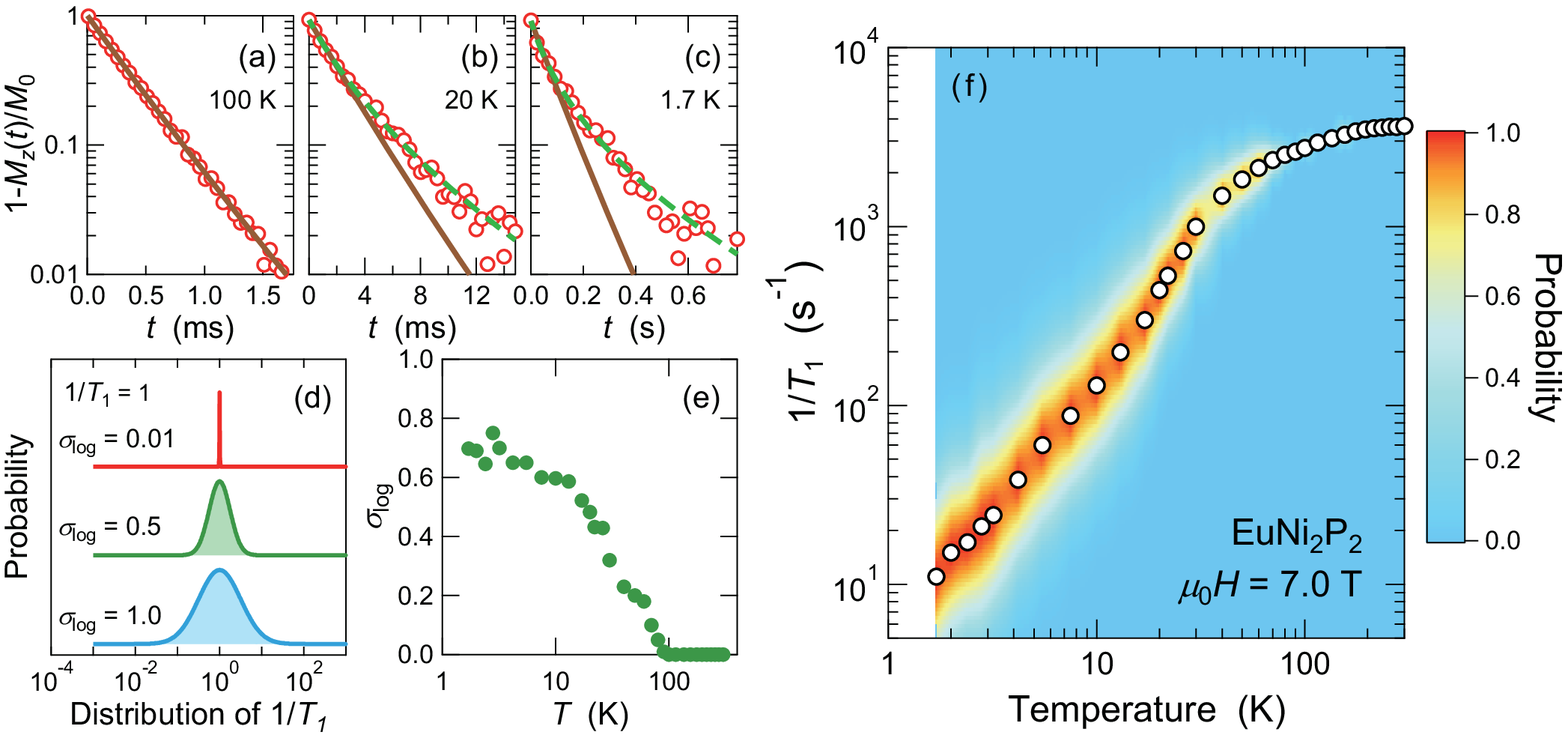}
  \end{center}
  \caption{(color online) (a), (b), and (c) $^{31}$P nuclear magnetization recovery curves at $T=100$, 20, and 1.7 K, respectively. The solid lines are a fit assuming a single $T_{1}$. The dashed lines are a fit assuming the Gaussian distribution of $T_{1}$ on a logarithmic scale. (d) Gaussian probability distribution of $1/T_{1}$ in the case of various $\sigma_{\rm log}$. (e) $T$ dependence of $\sigma_{\rm log}$ at $\mu_0H=7.0$ T, showing a distribution of $1/T_{1}$ below $\sim$80 K. (f) $T$ dependence of $1/T_{1}$ at $\mu_0H=7.0$ T. A distribution of $1/T_{1}$ is displayed by a color plot.}
  \label{fig3}
\end{fullfigure}

To investigate the electronic state furthermore, the nuclear magnetization recovery curve was measured at the peak frequency of the $^{31}$P NMR spectrum.
%更に詳細に調べるため、核スピン-格子緩和時間T1の測定を行った。
%NMRスペクトルのピーク周波数で測定を行った。
When an electronic state is homogeneous, the recovery curve for $I=1/2$ is generally described as a single exponential function given by
\begin{equation}
1-\frac{M_{z}(t)}{M_0} = e^{-\frac{t}{T_1}},\label{ExpRecov}
\end{equation}
where $M_{0}$ and $M_{z}(t)$ are the nuclear magnetization at the thermal equilibrium condition and the nuclear magnetization at a time $t$ after the saturation pulse, respectively.
Figures 3(a)--3(c) show the recovery curve measured at temperatures of 1.7, 20, and 100 K, with $\mu_0H = 7.0$ T.
Above 100 K, we can fit the data well by Eq. (\ref{ExpRecov}) as indicated by the solid line in Fig. 3(a) and determine $T_1$ uniquely.
However, the recovery curve begins to deviate from the single exponential behavior below about 80 K.
This suggests that the electronic state becomes inhomogeneous below about 80 K for some reason.
In this case, the two-component fit is often employed.
The two-component fit means that two electronic states corresponding to each $T_1$ are mixed; however, it is unlikely to occur in EuNi$_2$P$_2$.
Instead, the analysis assuming the distribution of $T_1$ is more plausible.
The stretch exponential function $\exp \left[-\left( \frac{t}{T_1} \right)^\beta \right]$ is often used for the analysis of the recovery curve when $T_1$ is distributed.
Detailed studies of the value and distribution of $T_1$ of the stretch exponential function for the exponent $\beta$ were reported.\cite{Berberan2005,Johnston2006}
Since the distribution of $T_1$ changes with the exponent $\beta$, it is necessary to pay attention to the analysis using the stretch  exponential recovery function when $\beta$ shows a temperature variation.
%Fig. 3(a)-3(c)に$H=7.0$ Tで測定した$^{31}$P nuclear magnetization recovery curve を示す。
%%100K以上ではrecovery curveは実線で示すように上式でよくフィットする事ができ、ユニークにT1を決めることができた。
%しかしながら、100K以下で緩和曲線は曲がり始め上式ではフィットすることができなくなった。
%これは電子状態が何らかの理由により、不均一になっていることを示唆している。
%この様な場合、しばしば、two component fitによる解析が行われる。
%2成分フィットでは短いT1を持つ電子状態と長いT1を持つ電子状態の2相が混在する事を意味するが、この物質で2相分離の電子状態が実現しているとは考えがたい。
%それよりも、T1が有限の幅を持って分布している解析がより好ましいと考えられる。
%T1が分布する場合のフィッティング式の一つとしてストレッチexponential functionがある。
%ストレッチexp関数に関する詳細な研究がこれまでに行われており、exponentβと得られるT1の値、そしてその分布の様子について調べられている。
%T1の分布の大きさ（exponentβ）の値によりもっともらしいT1も変化するため、T1の分布の大きさが温度依存する場合にストレッチexpによる解析は注意を要する。
%On the other hand, Mitrovi\'{c} et al. reported that an analysis of the recovery curve by assuming the Gaussian distribution of $1/T_1$ on a logarithmic scale:\cite{Mitrovic2008}
On the other hand, an analysis assuming the Gaussian distribution of $1/T_1$ on a logarithmic scale has been reported by Mitrovi\'{c} et al.\cite{Mitrovic2008}
According to that, the recovery curve can be written as follows
%一方で、T1がログスケールでガウス分布するとして緩和曲線を解析する方法がMitrovi\'{c}らにより報告されている。
%その場合の緩和曲線は以下のようになる。
%\begin{equation}
%\mathcal{M}_G(t) = \frac{1}{\sigma_{\rm log}}\sqrt{\frac{2}{\pi}}\int e^{\frac{-2(\log_{10}R_1-\log_{10}T_1^{-1})^2}{\sigma_{\rm log}^2}}e^{-tR_1}d(\log_{10}R_1)
%\end{equation}
\begin{equation}
\mathcal{M}_G(t) = \int \mathcal{P}_{\sigma_{\rm log},W_1}(R_1) e^{-tR_1} d(\log_{10}R_1).\label{GaussRecov}
\end{equation}
Here, $\mathcal{P}_{\sigma_{\rm log},W_1}(R_1)$ describes the relaxation rate distribution given by
\begin{eqnarray}
\mathcal{P}_{\sigma_{\rm log},W_1}(R_1) &=& \frac{1}{\sigma_{\rm log}}\sqrt{\frac{2}{\pi}}e^{\frac{-2(\log_{10}R_1-\log_{10}W_1)^2}{\sigma_{\rm log}^2}},
%m(t,R_1) &=& e^{-tR_1}
\end{eqnarray}
where $\sigma_{\rm log}$ is the width of the distribution on a $\log_{10}$ scale and $W_1$ is the center of the Gaussian.
Therefore, $W_1$ is considered as the most plausible relaxation rate. 
As an example, the distributions of the relaxation rate calculated with $W_1 = 1$ and various $\sigma_{\rm log}$ are shown in Fig. 3(d).
In the limit of $\sigma_{\rm log}\to 0$, Eq. (\ref{GaussRecov}) becomes single exponential, resulting in Eq. (\ref{ExpRecov}).
Therefore, we regard $W_1$ as $1/T_1$ in the following discussion.
The dashed lines in Figs. 3(b) and 3(c) are fit using Eq. (\ref{GaussRecov}), which reproduces the experimental data well.
We analyzed the recovery curve at each temperature using Eq. (\ref{GaussRecov}), and obtained $T_1$ and its distribution $\sigma_{\rm log}$.
The distribution of $1/T_1$ below 80 K increases with decreasing temperature as seen in Fig. 3(e).
%式5を用いてデータをフィットした結果がFig. 3(b)および3(c)の破線であり、実験データをよく再現している。
%我々はそれぞれの温度での回復曲線について式5を用いた解析を行い、T1とその分布$\sigma_{\rm log}$を求めた。
%Fig. 3(e)に示すように、100K以下で$1/T_1$の分布が増大して行っている様子が分かる。

The temperature dependence of $1/T_1$ at $\mu_0H=7.0$ T obtained by the analysis mentioned above is shown in Fig. 3(f).
The distribution of $1/T_1$ is also displayed as a color plot.
$1/T_1$ shows an almost constant behavior at around 300 K and begins to decrease slightly below about 200 K.
A further distinct decrease in $1/T_1$ is observed below 40 K.
At the lowest temperature, $1/T_1$ is almost proportional to temperature, indicating that the Fermi-liquid state with heavy quasiparticles is realized.

%以上の解析により求めた7Tで測定した1/T1の温度依存性をFig. 3(f)に示す。
%$1/T_1$の分布についての様子はカラープロットとして示している。
%300 K付近では$1/T_1$はほぼ一定値を示すが、200 K以下で$1/T_1$はわずかに減少し始める。
%$\sim$40 K以下ではより顕著な$1/T_1$の減少が観測された。
%より低温では$1/T_1$がほぼ温度に比例することから、重い準粒子が形成されたフェルミ液体状態が実現していることが示された。

\begin{figure}[tb]
  \begin{center}
    \includegraphics[keepaspectratio=true,width=80mm]{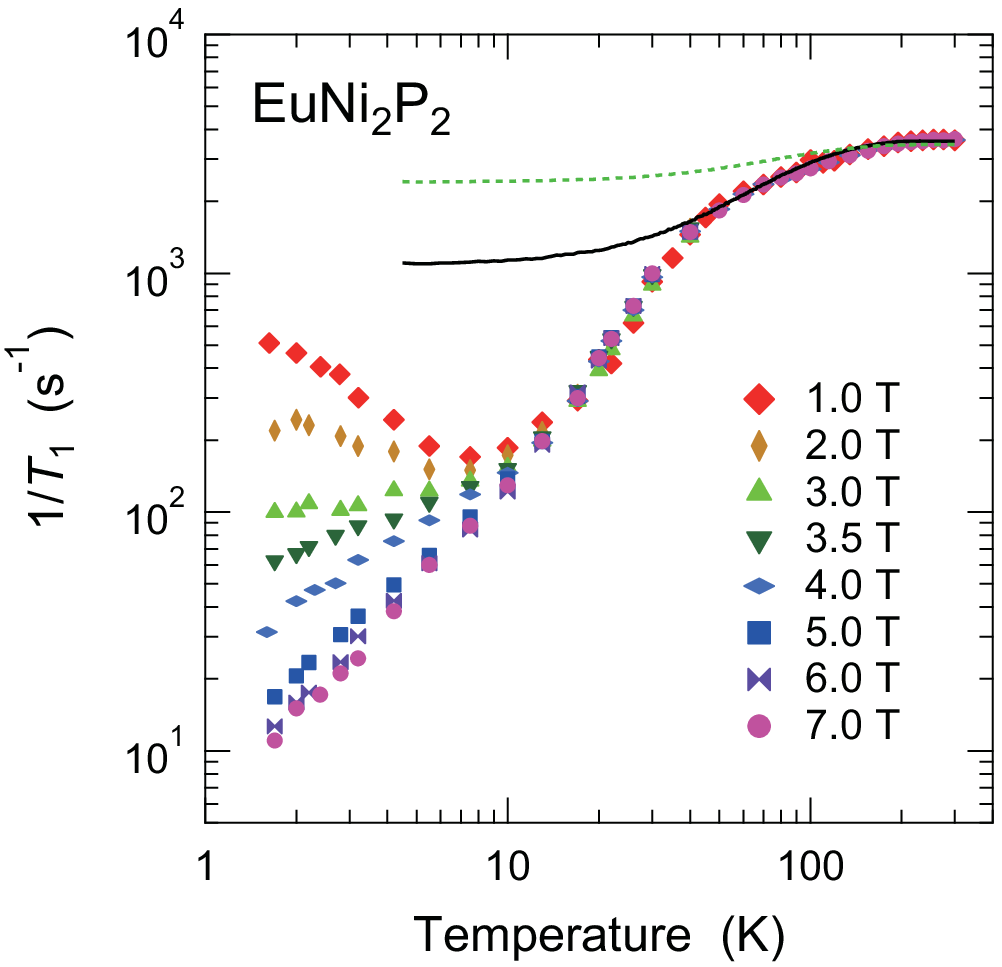}
  \end{center}
  \caption{(Color online) $T$ dependence of $1/T_{1}$ under various external magnetic fields. The dotted and solid lines indicate calculation results assuming that $1/T_{1}$ is related to the average valence of Eu.}
  \label{fig4}
\end{figure}
For further investigation, the magnetic field dependence of $1/T_1$ was measured.
In all the measured magnetic fields, the recovery curves below 80 K deviate from the single exponential behavior.
The temperature dependence of $1/T_1$ obtained using Eq. (\ref{GaussRecov}) in various magnetic fields is shown in Fig. 4.
In the high-temperature region above 15 K, $1/T_1$ is not affected by the magnetic field.
On the other hand, a significant magnetic field dependence was observed below 15 K, revealing the development of the spin fluctuations in low-temperature and low-magnetic-field regions.
%さらに詳細に調べるために$1/T_1$の磁場依存性について測定を行った。
%測定したすべての磁場において、緩和曲線のsingle exponential からのずれが約80K以下で観測された。
%よって、Eq.5を用いて、$1/T_1$を求めた。
%様々な磁場における$1/T_1$の温度依存性をFig. 4に示す。
%高温領域では磁場による影響は全くない。
%一方、約10K以下では顕著な磁場依存性が観測された。
%この結果は、低温低磁場領域において、スピン揺らぎが発達していることを示している。

\section{Discussion}

We discuss the relationship between the $4f$ electron state and $1/T_1$.
As described above, $1/T_1$ is almost constant at around 300 K.
When the nuclear magnetic relaxation is caused by fluctuating local moment, $1/T_1$ is given by \cite{Moriya1956}
\begin{equation}
\frac{1}{T_1} = \sqrt{2\pi}(g\gamma_n A_{\rm hf})^2\frac{J(J+1)}{3\omega_{\rm fl}},\label{localT1}
\end{equation}
where $\omega_{\rm fl}$ is the local moment fluctuation frequency, assumed as $\omega_{\rm fl} \gg \gamma_n H$.
If a fluctuation due to the exchange interaction between localized moments is dominant, $\omega_{\rm fl}$ is given as $\omega_{\rm ex}$ as
\begin{equation}
\omega_{\rm ex} = \frac{k_{\rm B}\theta_{\rm CW}}{\hbar}\sqrt{\frac{6}{zJ(J+1)}}.\label{omega-ex}
\end{equation}
Here, $z$ ($=8$) is the number of nearest-neighbor moments.
From the values of $\theta_{\rm CW} = -121.8$ K and $A_{\rm hf} = -6.53$ kOe/$\mu_{\rm B}$ obtained from the results of magnetic susceptibility and the Knight shift, the relaxation rate by the local moment fluctuation was estimated as $1/T_{1}\sim 8\times 10^{4}$ s$^{-1}$ for $J = 7/2$ of Eu$^{2+}$.
The calculated $1/T_1$ is one order of magnitude larger than the experimental value, suggesting that there is another fluctuation that is different from the fluctuation derived from the exchange interaction.
%The calculated $1/T_1$ is one order of magnitude larger than the experimental value, suggesting that there is a fluctuation other than local moment fluctuation.
In the case of EuNi$_2$P$_2$, the valence of Eu is reported to be Eu$^{2.25+}$ at 300 K.\cite{Nagarajan1985}
Therefore, the valence fluctuations between Eu$^{2+}$ and Eu$^{3+}$ affect the local moment fluctuation, and $\omega_{\rm fl}$ may be expressed as $\omega_{\rm fl} = \omega_{\rm ex} + \omega_{\rm vf}$.
%Therefore, the valence fluctuation between Eu$^{2+}$ and Eu$^{3+}$ may affect the local moment fluctuation.
%Considering the valence fluctuation, the local moment fluctuation frequency may be expressed as $\omega_{\rm fl} = \omega_{\rm ex} + \omega_{\rm vf}$.
Here, $\omega_{\rm vf}$ is a frequency of the valence fluctuation.
The measured $1/T_1$ being smaller than the calculated value that takes $\omega_{\rm ex}$ into consideration indicates the presence of valence fluctuations in EuNi$_2$P$_2$.
%These results indicate that EuNi$_2$P$_2$ has not only magnetic fluctuations but also valence fluctuations.
%Therefore, the valence fluctuation increases $\omega_{\rm fl}$ and $1/T_1$ decreases as observed.
%磁化率とナイトシフトの測定から、$\theta_{\rm CW} = -121.8$ Kと$A_{\rm hf} = 6.528$ kOe/$\mu_{\rm B}$と求められている。
%Eu$^{2+}$として、$J=7/2$とすると、$1/T_{1}\sim 8\times 10^{4}$ s$^{-1}$が得られた。
%この値は実験値と比べて一桁大きい値となっており、local moment fluctuation以外の揺らぎが存在することを示唆している。
%EuNi2P2のEuの価数は300Kにおいても00価と報告され、中間価数状態を取る。
%よってEu2+とEu3+の間の価数揺らぎが効いていると考えられる。
%価数揺らぎの周波数を$\omega_{\rm vf}$とすると、local moment fluctuationは簡単には$\omega_{\rm fl} \sim \omega_{\rm ex} + \omega_{\rm vf}$と書け、2価の場合よりも増大する。
%その結果、計算から求められた$1/T_1$が実験値と比べて1桁小さくなっていると考えられる。
%よって、1/T1の測定値がWEXを考慮に入れた計算値と比べて小さいことは、EuNi2P2における価数揺らぎの存在を示している。

According to Eq. (\ref{localT1}), $1/T_1$ is constant regardless of temperature; however, the measured $1/T_1$ depends on temperature even in the high-temperature region as shown in Fig. 4.
As described above, the $K-\chi$ plot shows a linear behavior, i.e., the hyperfine coupling constant does not change above 40 K.
Therefore, the decrease in $1/T_1$ below 200 K reflects the change in the $4f$ electron state of Eu.
It is reported from a M\"{o}ssbauer experiment that the average valence of Eu in EuNi$_2$P$_2$ is changed to the nonmagnetic side with decreasing temperature.\cite{Nagarajan1985}
Hence, the temperature variation in $1/T_1$ near 200 K is presumably related to the average valence of Eu.
Assuming that the nuclear magnetic relaxation by the local moment fluctuation of the $4f$ electron is derived from the Eu$^{2+}$ state, $1/T_1^{\rm fl}$ can be written as
%$1/T_1$がEq. (\ref{localT1})に従うとすると、$1/T_1$は温度依存をしないが、実験データは温度減少に従い、わずかに減少している。
%$K-\chi$プロットから40 K以上ではhyperfine coupling constantに変化はないため、Euの4f電子状態の変化を反映していると考えられる。
%メスバウアー測定からEuNi2P2におけるEuの平均価数は温度変化を示し、温度降下に伴い増加、つまり非磁性側に変化することが示されている。
%よって、高温における$1/T_1$の温度変化は平均価数と関係していると考えられる。
%Eu$^{2+}$の割合により$1/T_1$が変化すると仮定すると、平均価数$V_{\rm av}$との最も簡単な関係は以下のようになるだろう
\begin{equation}
\frac{1}{T_1^{\rm fl}} = a(b-V_{\rm av}).
\label{lotalT1forAV}
\end{equation}
Here, $V_{\rm av}$ is the average valence of Eu, and $a$ and $b$ are fitting parameters.
If $1/T_1^{\rm fl}=0$ for the trivalent Eu state, $b = 3$.
The dotted line in Fig. 4 is the calculated result with $b = 3$, which does not reproduce the measured $1/T_1$ well.
In contrast,  $1/T_1$ is well reproduced by $b = 2.58$ in a wide temperature range as shown by the solid line in Fig. 4.
This indicates that the intermediate valence state is not a simple combination of divalent and trivalent states, and that the valence fluctuation plays an important role in low-energy spin fluctuations.
%This indicates that the intermediate valence state is not a simple combination of the divalent and the trivalent states.

A remarkable decrease in $1/T_1$, compared with the estimated one from the temperature variation in the Eu average valence, was observed below $T^{\ast }\simeq 40$ K.
This indicates that the shielding of the local moment due to conduction electrons, that is, the Kondo effect, is caused.
The characteristic temperature $T^{\ast }$ is close to the Kondo temperature $T_{\rm K} \sim 80$ K determined from the electrical resistivity.\cite{Hiranaka2013}
%The temperature at which $1/T_1$ begins to distribute is close to $T_{\rm K}$ as shown in Fig. 3(e)-(f); therefore, we speculate that the distribution of $1/T_1$ is due to the distribution of $T_{\rm K}$.
The temperature at which $1/T_1$ begins to distribute is close to $T_{\rm K}$ as shown in Figs. 3(e) and 3(f); therefore, the distribution of $1/T_1$ is considered to be associated with the distribution of $T_{\rm K}$, which is presumed to be due to a slight inhomogeneity of the average valence.
Since $1/T_1$ does not show a remarkable magnetic field dependence above 15 K, it is apparent that the valence state and Kondo effect are not affected by the magnetic field of about 7 T.
%Euが3価の場合$1/T_1^{\rm fl}=0$とすると、$b=3$となる。
%その場合のfitting resultがFig. 4のdotted lineである。
%あまり実験結果を上手く再現していない。
%Fig. 4のsolid lineに示すように、$b = 000$とすると広い温度範囲でフィットすることができた。
%これは中間価数状態というのは単純に3価と2価の足し合わせの電子状態ではないことを示している。
%価数の変化を考慮した実線よりも急激な減少が40K以下で見られる。
%これは近藤効果による局在モーメントと伝導電子の混成による準粒子を形成に伴う電子状態の変化による物だと考えられる。
%図3のカラープロットに示すように、1/T1が分布し始める温度は、1/T1の顕著な減少を示す温度に近い。
%よって、我々は1/T1の分布は近藤温度TKの分布による物だと推測する。
%15 K以上では$1/T_1$は顕著な磁場依存性を示さないことから、高温での価数状態や磁気状態、近藤温度は磁場による影響はない。

Next, we focus on the temperature range below 10 K.
A clear magnetic field dependence was observed as shown in Fig. 4.
An increase in $1/T_1$ accompanied by the development of spin fluctuation was observed at $\mu_0H = 1.0$ T.
An increase in the magnetic field suppresses the enhancement of $1/T_1$, and the $T_1T = const.$ behavior inherent to the Fermi-liquid state is realized at $\mu_0H = 7.0$ T.
Fisher et al. also observed an anomaly at low temperatures from the specific heat measurements.\cite{Fisher1995}
The specific heat divided by temperature, $C/T$, shows an upturn below 5 K and is suppressed by the application of a magnetic field of 7 T.
This behavior is similar to the magnetic field dependence of $1/T_1$.
However, it has been pointed out that the anomaly in the low-temperature specific heat should be related to a small impurity contained in the sample,\cite{Fisher1995} thus suggesting that the observed spin fluctuation in $1/T_1$ is also caused by impurities.
Generally, the nuclear magnetic relaxation by impurities is easily suppressed by a small magnetic field.
In EuNi$_2$P$_2$, the magnetic field of about 7 T, as seen in Fig. 4, is required to suppress the spin fluctuation at low temperatures.
Therefore, it is unclear whether the anomaly observed in $1/T_1$ is caused by magnetic impurities or it is a novel phenomenon unique to the heavy electron state of the Eu compound with the intermediate valence state.

\begin{figure}[tb]
  \begin{center}
    \includegraphics[keepaspectratio=true,width=70mm]{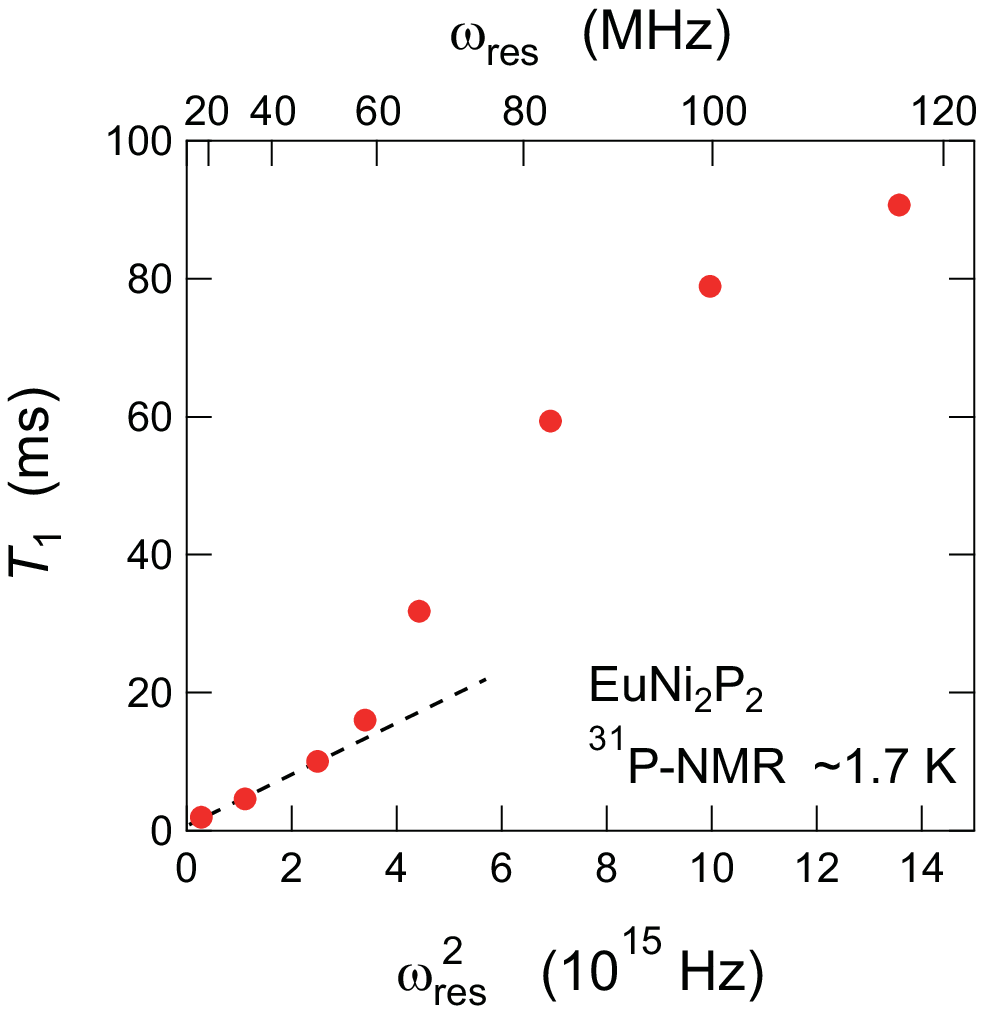}
  \end{center}
  \caption{(Color online) $\omega_{\rm res}^2$ dependence of $T_1$ at the lowest temperature. As indicated by the dashed line, a deviation from the linear relation occurs at more than about 50  MHz ($\mu_0H \sim 3$ T).}
  \label{fig5}
\end{figure}
In addition, the influence of the nuclear magnetism of the Eu nucleus is also conceivable.
For example, in PrFe$_4$P$_{12}$, relaxation by the nuclear magnetism of $^{141}$Pr nuclei with a nuclear magnetic moment of 4.14$\mu_{\rm N}$ has been observed.\cite{Ishida2005}
The $1/T_1$ of PrFe$_4$P$_{12}$ in the low-temperature and low-magnetic-field regions showed a decrease with increasing magnetic field (resonant frequency $\omega_{\rm res}$).
For the magnetic field dependence of $1/T_1$, the fluctuation of the $^{141}$Pr nuclear magnetic moment was analyzed, and it was found that $T_1$ and $\omega_{\rm res}$ follow the relation $T_1=A+B\omega_{\rm res}^2$.
Here, $A$ and $B$ are parameters related to the characteristic frequency and magnitude of the fluctuation of the nuclear magnetic moment.
From this result, it was concluded that the magnetic field dependence of $1/T_1$ of PrFe$_4$P$_{12}$ is due to the magnetism of $^{141}$Pr nuclei.
The linear relation between $T_1$ and $\omega_{\rm res}^2$ was also found in $^{17}$O NMR in NpO$_2$, which was explained by the cross-relaxation process by $^{237}$Np nuclear spin.\cite{Tokunaga2006}
%The linear relation between $T_1$ and $\omega_{\rm res}^2$ has also been found in $^{17}$O NMR in NpO$_2$ and is explained as due to the cross-relaxation process with $^{237}$Np nuclear spin.\cite{Tokunaga2006}
Eu has two isotopes, $^{151}$Eu and $^{153}$Eu, and their nuclear magnetic moments are 3.47$\mu_{\rm N}$ and 1.53$\mu_{\rm N}$, respectively.
Therefore, the relaxation of the $^{31}$P nucleus due to the fluctuation of the Eu nuclear dipole magnetic field may have been observed.
To verify these, we plotted the $\omega_{\rm res}^2$ variation of $T_1$ at the lowest temperature as shown in Fig. 5.
As is apparent from the dashed line in Fig. 5, a linear relation was seen only in a very narrow range of $\omega_{\rm res} < 50$ MHz ($\mu_0H \sim 3$ T).
Although we cannot eliminate the possibility of Eu nuclear magnetism, we believe that the increase in $1/T_1$ is caused by characteristic fluctuations of other origins.
To elucidate these points, further studies including those of other heavy-fermion Eu compounds are required.
Theoretical studies of the intermediate valence state and $c-f$ hybridization in Eu compounds are also expected.
%次に10K以下の低温領域に着目する。
%Fig. 4から明らかな様に、明瞭な磁場依存性が観測された。
%本研究での最低磁場である$\mu_0H = 1.0$ Tでは温度降下に従い$1/T_1$は増大していった。
%これはスピン揺らぎが低温で発達していることを意味している。
%外部磁場の増大に従い、スピン揺らぎは抑制され、$\mu_0H = 7.0$ Tではフェルミ液体状態を示す$(T_1T)^{-1}=const.$が観測された。
%この様な低温での磁場依存性は、比熱にも観測されている。
%Fisherらによる極低温領域までの比熱測定によると、ゼロ磁場の$C/T$にわずかなup turnが観測されている。
%C/Tは5K以下で上昇を示し、7Tの磁場の印加により抑制される。
%これは1/T1の振る舞いと類似している。
%しかしながら、この低温比熱における異常は試料に含まれるわずかな不純物による可能性が指摘されており、よって、観測されたスピン揺らぎも不純物による可能性が示唆される。
%通常の磁性不純物による緩和は磁場で容易に抑制される。
%しかしながら、今回観測された低温でのスピン揺らぎの抑制には数Tの磁場が必要であり、これが本当に不純物によるものかはunknownである。
%上記以外にも、Eu核磁性による影響も考えられる。
%例えば、PrFe4P12において、141Pr核の核磁性による緩和が観測されている。
%低温および低磁場領域におけるPrFe4P12の1/T1は磁場(共振周波数)の増加に伴って減少を示した。
%この緩和率の磁場依存性について141Pr核磁気モーメントの揺らぎに基づいた解析が行われ、T1と測定周波数がT1=A+Bw^2の関係に従うことが見いだされた。
%ここで、AとBは核磁気モーメントの揺らぎの特性周波数と大きさに関連したパラメータである。
%この結果から、PrFe4P12の1/T1の磁場依存性は4muNの大きさの核磁気モーメントを持つ141Pr核の磁性によるものだと結論された。
%T1とw^2の間の線形関係はNpO2における17O-NMRでも見いだされており、237Np核スピンによるcross-relaxationプロセスに因ると説明されている。
%The cross-relaxation process, driven by the 237Np spins, is conveyed to the 17O via a greatly enhanced indirect nuclear spin-spin coupling.

%Euは二つの同位体151Euと153Euをもち、それぞれの核磁気モーメントはmuEu1とmuEu2となる。
%よって、Eu核双極子磁場の揺らぎによる31P核の縦緩和が観測されている可能性がある。
%これらについて検証するために、図5の様に最低温度でのT1のWres2変化をプロットした。
%図5の破線から明らかなように、非常に狭い範囲のみで線形関係が見られることが分かる。
%Eu核磁性の可能性を完全に排除することはできないが、我々は他の原因による特異な揺らぎにより1/T1の増大が生じていると考える。
%更なる研究が必要であると考えられる。
%更に、中間価数状態を取るEu化合物におけるf電子状態に関する理論的な研究も期待される。

\section{Summary}
To clarify the electronic state of the intermediate valence compound EuNi$_2$P$_2$ from the microscopic viewpoint, $^{31}$P NMR measurements were carried out in various magnetic fields up to 7 T.
In the high-temperature region, the Knight shift follows the Curie--Weiss law, indicating that the $4f$ electrons of Eu are localized.
The localized $4f$ electronic state was also observed in $1/T_1$, and the temperature variation in $1/T_1$ corresponding to the change in the average valence of Eu was found.
Below about 40 K, $1/T_1$ showed an apparent decrease with decreasing temperature.
In addition, the absolute value of the Knight shift showed the maximum at approximately 40 K, and it became almost constant at low temperatures.
These results reveal the occurrence of the Kondo effect in EuNi$_{2}$P$_{2}$.
%The characteristic point is that 
$1/T_1$ exhibits a clear magnetic field dependence at low temperatures, which is attributed to the valence fluctuation caused by the intermediate valence state of EuNi$_2$P$_2$.

%中間価数状態をとるEu化合物EuNi2P2について微視的な視点から電子状態を明らかにするため、P NMR測定を行った。
%高温領域において、ナイトシフトはキュリーワイス則に従う振る舞いを示し、4f電子は局在的である事が示された。
%局在的な4f電子状態はまた、1/T1においても観測され、さらに、Euの平均価数の変化に対応した1/Tの温度変化が見いだされた。
%40K以下で1/T1は明瞭な減少を示した。
%また、ナイトシフトについても40K以下でその絶対値は最大値を取り、最低温ではほぼ一定となった。
%これらの結果から、磁気的な近藤効果が生じていることが明らかにされた。
%特徴的な点は、低温において1/T1が明瞭な磁場依存性を示すことで、これはEuNi2P2の中間価数状態に起因した価数揺らぎに因るものだと推測される。

\section*{Acknowledgments}
We would like to thank K. Magishi, Y. Tokunaga, and K. Ishida for useful comments and discussion.
This work was supported by JSPS KAKENHI Grant Numbers JP20102002, JP23540418, JP16K05453, JP17K05547 and JGC-S SCHOLARSHIP FOUNDATION.

\end{document}